\documentstyle[12pt]{article}
\begin{document}

\begin{flushright}
August 2000

OU-HET 360
\end{flushright}

\begin{center}

\vspace{5cm}
{\Large BPS D-branes after Tachyon Condensation} 

\vspace{2cm}
Takao Suyama \footnote{e-mail address : suyama@funpth.phys.sci.osaka-u.ac.jp}

\vspace{1cm}

{\it Department of Physics, Graduate School of Science, Osaka University, }

{\it Toyonaka, Osaka, 560-0043, Japan}

\vspace{4cm}

{\bf Abstract} 

\end{center}

We construct an effective action describing brane-antibrane system containing $N$ D-branes 
and $N$ $\bar{\mbox{D}}$-branes. 
BPS equations for remaining D-branes after tachyon condensation are derived and their 
properties are investigated. 
The value of the D-brane tension and the number of brane bound states are discussed. 

\newpage

{\bf {\large 1. Introduction}}

\vspace{5mm}

The importance of tachyon in string theory has been shown in recent researches. 
The presence of the tachyon indicates the instability of brane-antibrane systems 
\cite{tachyon} and this system decays into a vacuum \cite{nonBPS}. 
This is explained that the tension of the brane-antibrane system is cancelled 
by the minimum value of the tachyon potential $V(T)$, and the occurance of such 
cancellation is examined by using string field theory \cite{SFT}. 
The study of the tachyon condensation also provides new objects in string theory; 
non-BPS D-branes \cite{nonBPS}. 
To classify these objects as well as the ordinary BPS D-branes K-theory is introduced into 
string theory and has been intensively studied \cite{Ktheory}. 

In our previous paper \cite{TS}, we discussed BPS D-branes after tachyon condensation in 
terms of a spacetime effective theory. 
D-branes were described as classical solutions of some BPS equations, and such description 
could deal with both paralell branes and intersecting branes. 
In addition, the form of the tachyon potential $V(T)$ was determined by requiring the 
stability of the classical solutions. 

We will generalize in this paper the effective theory of the brane-antibrane system to the 
one which describes the system containing $N$ Dp-branes and $N$ $\bar{\mbox{D}}$p-branes. 
This is a non-Abelian gauge theory coupled to the tachyon field, and this theory captures 
several properties of the brane system which cannot be represented by our old one. 

This paper is organized as follows. 
We briefly review our previous work in section 2. 
The action of the multiple brane-antibrane system is constructed and the supersymmetry 
of this theory is discussed in section 3. 
The properties of the solutions of BPS equations and the fluctuations around them are 
investigated in section 4, and the asymptotic form of the tachyon field is determined in 
section 5. 
Section 6 is devoted to discussions. 

\vspace{1cm}

{\bf {\large 2. BPS branes after tachyon condensation}}

\vspace{5mm}

The brane-antibrane system has shown to be unstable. 
The instability is due to the tachyonic mode coming from the string stretched between brane 
and antibrane. 
The calculations using string field theory show that the tension of the system is 
completely cancelled after tachyon condensation \cite{nonBPS}. 
This means that the brane-antibrane system will decay into some stable state. 
The resulting stable configuration is either a vacuum or lower dimensional branes, and 
the latter can be classified by K-theory \cite{Ktheory}. 

In our previous paper \cite{TS}, we discussed the BPS D-branes which remain after tachyon 
condensation by using a spacetime action. 
The action of D9-$\bar{\mbox{D}}$9 system is the following,
\begin{equation}
S = \frac1{g^2}\int d^{10}x \Bigl\{ -\frac14 F^{(r)}_{\mu\nu}F^{(r)\mu\nu}+\frac i2 
      \bar{\psi}^{(r)}
      \Gamma^{\mu}\partial_{\mu}\psi^{(r)} -|D_{\mu}T|^2-V(T) \Bigr\} , \label{U1action}
\end{equation}
where $D_{\mu}T=\partial_{\mu}T-iA^{(r)}_{\mu}T$. 
$A^{(r)}_{\mu}$ and $\psi^{(r)}$ are ten-dimensional $U(1)$ gauge multiplet on the 
brane-antibrane system (so called relative $U(1)$), and $T$ is the tachyon field. 
There also exists a charged massless fermion which is ignored here because it is irrelevant 
for the following discussions. 
The action of Dp-$\bar{\mbox{D}}$p system can be obtained from (\ref{U1action}) by 
dimensional reduction. 

According to the arguments in string theory, this system should contain a stable solution 
which preserves some of the supersymmetries although the action (\ref{U1action}) is not 
supersymmetric. 
Now let us consider the following ``supersymmetry'' transformations,
\begin{eqnarray}
&& \delta A^{(r)}_\mu = \frac i2 \bar{\epsilon}\,\Gamma_{\mu}\psi \\
&& \delta \psi^{(r)} = -\frac14 F^{(r)}_{\mu\nu}\Gamma^{\mu\nu}\epsilon 
                     -\frac12 (|T|^2-\zeta)\Gamma^1\Gamma^2\epsilon \\
&& \delta T = 0
\end{eqnarray}
where $\zeta>0$. 
The action $S$ is not invariant under the above transformations, and one of the conditions 
for $\delta S=0$ is 
\begin{equation}
D_1T+iD_2T=0. \label{NOeq1}
\end{equation}
$\delta \psi=0$ for general $\epsilon$ means 
\begin{equation}
F^{(r)}_{12} +|T|^2-\zeta = 0. \label{NOeq2}
\end{equation}
Thus we conclude that the solutions of (\ref{NOeq1}) and (\ref{NOeq2}) are the BPS 
solutions preserving 16 supercharges. 
(\ref{NOeq1})(\ref{NOeq2}) are the Nielsen-Olesen vortex equations \cite{vortex} and the 
solutions can be interpreted as BPS D-branes of codimension two. 

\vspace{1cm}

{\bf {\large 3. Non-Abelian generalization}}

\vspace{5mm}

In this section we will construct an effective action of the brane-antibrane system which 
consists of $N$ D9-branes and $N$ $\bar{\mbox{D}}$9-branes. 
This system is described by a gauge theory with gauge group $SU(N)\times SU(N)\times U(1)$. 
The relevant fields on the branes are the followings: 
$A^{(1)}_\mu,\psi^{(1)}$ are $SU(N)$ gauge multiplet on D9-branes and $A^{(2)}_\mu,\psi^{(2)}$ 
are $SU(N)$ gauge multiplet on $\bar{\mbox{D}}$9-branes.  
$A^{(r)}_\mu,\psi^{(r)}$ are the relative $U(1)$ multiplet. 
$T$ is the tachyon filed in $({\bf \mbox{N}},{\bf \bar{\mbox{N}}})$ representation of 
$SU(N)\times SU(N)$ and this has the unit $U(1)$ charge. 

The action of this system is
\begin{eqnarray}
&&S=\frac1{g^2}\int d^{10}x \Bigl[ -\frac14\mbox{tr} F^{(1)}_{\mu\nu}F^{(1)\mu\nu}
     +\frac i2\mbox{tr} \bar{\psi}^{(1)}\Gamma^{\mu}D_{\mu}\psi^{(1)}
     -\frac14\mbox{tr} F^{(2)}_{\mu\nu}F^{(2)\mu\nu} \nonumber \\
&&\hspace*{10mm}+\frac i2\mbox{tr}\bar{\psi}^{(2)}\Gamma^{\mu}D_{\mu}\psi^{(2)} 
     -\frac14F^{(r)}_{\mu\nu}F^{(r)\mu\nu}
     +\frac i2\bar{\psi}^{(r)}\Gamma^{\mu}\partial_{\mu}\psi^{(r)}
     -\mbox{tr}D_{\mu}T^{\dagger}D^{\mu}T -V(T) \Bigr]. \nonumber \\ \label{action}
\end{eqnarray}
The field strengths and the covariant derivatives are defined as follows. 
\begin{eqnarray}
&& F^{(\alpha)}_{\mu\nu} = \partial_{\mu}A^{(\alpha)}_{\nu}-\partial_{\nu}A^{(\alpha)}_{\mu}
                         -i[A^{(\alpha)}_\mu,A^{(\alpha)}_\nu] \hspace{10mm} (\alpha=1,2) \\
&& F^{(r)}_{\mu\nu} = \partial_{\mu}A^{(r)}_{\nu}-\partial_{\nu}A^{(r)}_{\mu} \\
&& D_\mu \psi^{(\alpha)} = \partial_\mu\psi^{(\alpha)}-i[A^{(\alpha)}_\mu,\psi^{(\alpha)}] \\
&& D_\mu T = \partial_\mu T-iA^{(1)}_\mu T+iTA^{(2)}_\mu
            -\frac i{\sqrt{N}}A^{(r)}_\mu T
\end{eqnarray}
The action (\ref{action}) is completely determined by the gauge invariance and the Lorentz 
invariance except for the tachyon potential $V(T)$. 
As in the $N=1$ case, there exists a massless fermion in 
$({\bf \mbox{N}},{\bf \bar{\mbox{N}}})$ representation of  $SU(N)\times SU(N)$ and the unit 
$U(1)$ charge, which is also ignored as in our previous paper. 
We have used the following normalization of the generators $t_a$ of  $SU(N)$.
\begin{equation}
\mbox{tr} \ t_a t_b = \delta_{ab}
\end{equation}

Then let us consider the ``supersymmetry'' transformations. 
\begin{eqnarray}
&& \delta A^{(\alpha)}_\mu = \frac i2 \bar{\epsilon}\,\Gamma_\mu\psi^{(\alpha)} 
                                \label{transf1} \\
&& \delta A^{(r)}_\mu = \frac i2 \bar{\epsilon}\,\Gamma_\mu\psi^{(r)} \\
&& \delta \psi^{(1)} = -\frac14F^{(1)}_{\mu\nu}\Gamma^{\mu\nu}\epsilon
                       -\frac12(TT^{\dagger}-\frac1N\mbox{tr} TT^{\dagger}\cdot 1)
                        \Gamma^1\Gamma^2\epsilon \\
&& \delta \psi^{(2)} = -\frac14F^{(2)}_{\mu\nu}\Gamma^{\mu\nu}\epsilon \\
&& \delta \psi^{(r)} = -\frac14F^{(r)}_{\mu\nu}\Gamma^{\mu\nu}\epsilon
                       -\frac1{2\sqrt{N}}(\mbox{tr} TT^{\dagger}-N\zeta)
                        \Gamma^1\Gamma^2\epsilon \\
&& \delta T = 0 \label{transf6}
\end{eqnarray}
The variation of the action (\ref{action}) is 
\begin{eqnarray}
&& \delta S = \frac1{g^2}\int d^{10}x \frac i2 \mbox{tr}\Bigl[
                D_\mu TT^\dagger\bar{\psi}^{(1)}
               (i\Gamma^\mu-\Gamma^\mu\Gamma^1\Gamma^2)\epsilon \nonumber \\
&& \hspace{10mm}-i T^\dagger D_\mu T\bar{\psi}^{(2)}\Gamma^\mu\epsilon 
                +\frac1{\sqrt{N}} D_\mu TT^\dagger\bar{\psi}^{(r)}
               (i\Gamma^\mu-\Gamma^\mu\Gamma^1\Gamma^2)\epsilon + h.c. \ \Bigr].
\end{eqnarray}
Therefore the conditions for $\delta S=0$ are
\begin{eqnarray}
&& D_1T+iD_2T=0 \label{condition1} \\
&& D_iT=0 \hspace{2cm} (i=0,3,\cdots,9) \\
&& T\psi^{(2)}=0. \label{condition3}
\end{eqnarray}
Under the conditions (\ref{condition1})-(\ref{condition3}), the action $S$ is invariant 
for general $\epsilon$. 
When we would like to discuss classical solutions which preserve only part of 16 
supercharges, the above conditions are relaxed in a suitable way.

\vspace{1cm}

{\bf {\large 4. Gauge symmetry on the D-branes}}

\vspace{5mm}

We will investigate in this section classical solutions which preserve 16 supercharges. 
In particular we will focus on the solutions corresponding to D-branes of codimension 
two. 
For such solutions we assume that they depend only on the coordinates $x^1$ and $x^2$, and 
$A^{(\alpha)}_{1,2},A^{(r)}_{1,2}$ are the only nonzero components of the gauge fields. 
Then $\delta \psi^{(\alpha)}=0,\delta\psi^{(r)}=0$ mean
\begin{eqnarray}
&& F^{(1)}_{12}+TT^\dagger-\frac1N\mbox{tr}TT^\dagger\cdot 1=0 \label{eq1} \\
&& F^{(2)}_{12}=0 \\
&& F^{(r)}_{12}+\frac1{\sqrt{N}}(\mbox{tr}TT^\dagger-N\zeta)=0. \label{eq2}
\end{eqnarray}
We set $A^{(2)}_\mu=0$ and define the $U(N)$ gauge fields 
\begin{equation}
A_\mu = A^{(1)}_\mu + \frac1{\sqrt{N}}A^{(r)}_\mu\cdot 1. \label{U(N)}
\end{equation}
Then the eqs.(\ref{eq1})(\ref{eq2}) become 
\begin{equation}
F_{12}+TT^\dagger-\zeta\cdot 1=0. \label{NAeq}
\end{equation}
This is a natural generalization of the Nielsen-Olesen vortex equation 
(\ref{NOeq1})(\ref{NOeq2}). 

The energy of the solutions of eqs.(\ref{condition1})(\ref{NAeq}) can be calculated as 
follows. 
\begin{eqnarray}
&& E=\frac1{g^2}\int d^{10}x \Bigl[ \frac12\mbox{tr}(F_{12}+TT^\dagger-\zeta\cdot 1)^2
    +\mbox{tr}|D_1T+iD_2T|^2+\zeta \mbox{tr}F_{12} \nonumber \\
&&\hspace{3cm} +V(T)-\frac12\mbox{tr}(TT^\dagger-\zeta\cdot 1)^2 \Bigr]
\end{eqnarray}
Thus the solutions are stable topologically if 
\begin{equation}
V(T)=\frac12\mbox{tr}(TT^\dagger-\zeta\cdot 1)^2
\end{equation}

Now we will consider the properties of the solutions of eqs.
(\ref{condition1})(\ref{NAeq}). 
For simpliciy, let $A_{1,2}$ and $T$ be diagonal matrices. 
\begin{eqnarray}
&& A_\mu=\mbox{diag}(A_{(1)\mu},\cdots,A_{(N)\mu}) \\
&& T=\mbox{diag}(T_{(1)},\cdots,T_{(N)}) 
\end{eqnarray}
Then eqs.(\ref{condition1})(\ref{NAeq}) become $N$ sets of the ordinary vortex equations 
(\ref{NOeq1})(\ref{NOeq2}). 
In this case the energy is 
\begin{equation}
E=\sum_{i=1}^N \frac{\zeta}{g^2}\int d^{10}x F_{(i)12}.
\end{equation}
The position of the vortex with respect to $F_{(i)12}$ is determined as the point 
where $T_{(i)}=0$. 
Therefore the number $n(x)$ of D-branes at some point $x$ is 
\begin{equation}
n(x)=N-\mbox{rank}(T(x)).
\end{equation}
This means that the solutions of eqs.(\ref{condition1})(\ref{NAeq}) can describe 
coincident D-branes as well as separated D-branes. 

Now we will discuss the fluctuations aruond the classical solutions. 
Suppose that $T$ has the form 
\begin{equation}
T=\mbox{diag}(0,\cdots,0,T_{n+1},\cdots,T_{N})
\end{equation}
at some point $x$. 
Consider the fluctuations of $A^{(2)}_\mu,\psi^{(2)},A^{(r)}_\mu,\psi^{(r)}$,
\begin{eqnarray}
&& A^{(2)}_\mu =A^{(2)c}_\mu +a^{(2)}_\mu \\
&& A^{(r)}_\mu =A^{(r)c}_\mu +a^{(r)}_\mu \\
&& \psi^{(2)}=\varphi^{(2)} \\
&& \psi^{(r)}=\varphi^{(r)} 
\end{eqnarray}
where the superscript $c$ indicates the classical solution of 
eqs.(\ref{condition1})(\ref{NAeq}). 
For the dynamics of the fluctuations to be supersymmetric, the conditions 
(\ref{condition1})-(\ref{condition3}) have to 
be satisfied, that is, 
\begin{eqnarray}
&& T(a^{(2)}_\mu +\frac1{\sqrt{N}}a^{(r)}_\mu \cdot 1)=0 \label{cond1} \\
&& T\varphi^{(2)}=0. \label{cond2}
\end{eqnarray}
Let us define the $U(N)$ gauge multiplet
\begin{eqnarray}
&& a_\mu = a^{(2)}_\mu +\frac1{\sqrt{N}}a^{(r)}_\mu \cdot 1 \\
&& \varphi=\varphi^{(2)}+\frac1{\sqrt{N}}\varphi\cdot 1.
\end{eqnarray}
Then the conditions (\ref{cond1})(\ref{cond2}) imply
\begin{equation}
a_\mu = \left( 
  \begin{array}{cc} 
     a_{n,\mu} & 0 \\ 0 & 0 
  \end{array} \right) \hspace{5mm} , \hspace{5mm}
\varphi=\left( 
  \begin{array}{cc}
     \varphi_n & 0 \\ 0 & 0 
  \end{array} \right),
\end{equation}
where $a_{n,\mu},\varphi_n $ are $n\times n$ hermitian matrices. 
This indicates that $U(n)$ gauge symmetry appears on the $n$ coincident D-branes. 
In fact, they are the fields in the super Yang-Mills theory with the ordinary supersymmetry 
transformations. 
If $T$ has the generic form, we have to set $a_\mu=0,\varphi=0$ to satisfy the conditions 
(\ref{cond1})(\ref{cond2}), thus the fluctuations are restricted on the branes. 

We have not mentioned the fluctuations $a^{(1)}_\mu$ of $A^{(1)}_\mu$. 
If we include them, the action contains a term $\mbox{tr}[A^{(1)c}_\mu,a^{(1)}_\nu]^2$. 
This indicates that in general $a^{(1)}_\mu$ become massive and can be ignored. 
However it may be possible for them to become massless, for example, at the brane 
intersection, if we consider such brane systems.

\newpage

{\bf {\large 5. Asymptotic behavior of the tachyon field}}

\vspace{5mm}

In this section we turn to the solutions which preserve part of 16 supercharges. 

First we will consider the solutions corresponding to intersecting branes. 
For example, there is an intersecting brane configuration which preserves 8 supercharges; 
two D7-branes extended along (01256789) and (03456789). 
The remaining supersymmetries correspond to the parameter $\epsilon$ satisfying 
\begin{equation}
\Gamma^1\Gamma^2\epsilon = \Gamma^3\Gamma^4\epsilon.
\end{equation}
In this case the conditions for $\delta S=0$ under the transformations 
(\ref{transf1})-(\ref{transf6}) are modified as follows.
\begin{eqnarray}
&& D_1T+iD_2T=0 \label{cond3} \\
&& D_3T+iD_4T=0 \label{cond4} \\
&& D_iT=0 \hspace{2cm} (i=0,5,\cdots,9) \\
&& T\psi^{(2)}=0
\end{eqnarray}

We assume, as in the previous section, that $A^{(2)}_\mu =0, A^{(1)}_i=0, A^{(r)}_i=0$ 
and the solutions depend only on the coordinates $x^1,x^2,x^3,x^4$. 
Then $\delta \psi^{(1)}=0,\delta\psi^{(r)}=0$ mean
\begin{eqnarray}
&& F^{(1)}_{12}+F^{(1)}_{34}+TT^\dagger-\frac1N\mbox{tr}TT^\dagger=0 \\
&& F^{(1)}_{13}-F^{(1)}_{24}=0 \\
&& F^{(1)}_{14}+F^{(1)}_{23}=0 \\
&& F^{(r)}_{12}+F^{(1)}_{34}+\frac1{\sqrt{N}}(\mbox{tr}TT^\dagger-N\zeta)=0 \\
&& F^{(r)}_{13}-F^{(r)}_{24}=0 \\
&& F^{(r)}_{14}+F^{(r)}_{23}=0.
\end{eqnarray}
In terms of the $U(N)$ gauge fields (\ref{U(N)}) the above equations can be rewritten as 
\begin{eqnarray}
&& F_{12}+F_{34}+TT^\dagger-\zeta\cdot 1=0 \label{Inteq1} \\
&& F_{13}-F_{24}=0 \\
&& F_{14}+F_{23}=0. \label{Inteq3}
\end{eqnarray}
Eqs.(\ref{cond3})(\ref{cond4})(\ref{Inteq1})-(\ref{Inteq3}) are the equations which 
describe the intersecting branes. 
One can show that the solutions of them are stable topologically when one takes 
$V(T)=\frac12\mbox{tr}(TT^\dagger-\zeta\cdot 1)^2$. 

We introduce the complex coordinates.
\begin{equation}
z^k=x^{2k-1}+ix^{2k} \hspace{1cm} (k=1,2)
\end{equation}
Then the BPS equations become as follows,
\begin{eqnarray}
& D_{\bar{a}}T=0 & \label{holo} \\
& F_{\bar{a}\bar{b}}=0 & \label{flat} \\
& -ig^{a\bar{b}}F_{a\bar{b}}+TT^\dagger-\zeta\cdot 1=0,& \label{nontri}
\end{eqnarray}
where $g_{a\bar{b}}$ is the metric in the complex coordinates; its nonzero components are 
$g_{z^1\bar{z}^1}=g_{z^2\bar{z}^2}=\frac12$. 

Similar equations can be derived for the solutions which preserve only 4 supercharges. 
Consider an intersecting brane configuration which contains D7-branes extended along 
(01234789),(03456789) and (01256789). 
The conditions for the remaining supersymmetries in this case are
\begin{equation}
\Gamma^1\Gamma^2\epsilon=\Gamma^3\Gamma^4\epsilon=\Gamma^5\Gamma^6\epsilon .
\end{equation}
The BPS equations are formally the same as that for D7-D7' system except that in this case 
$a,b$ take $z^1,z^2,z^3$. 

Eqs.(\ref{holo})(\ref{flat}) can be easily solved. 
From (\ref{flat}) $A_{\bar{a}}$ are determined to be
\begin{equation}
A_{\bar{a}}=iV\partial_{\bar{a}}V^{-1},
\end{equation}
where $V$ is an $N\times N$ complex matrix. 
Let $T=VT_0$. 
Then (\ref{holo}) becomes 
\begin{equation}
\partial_{\bar{a}}T_0=0.
\end{equation}
Thus $T_0$ is just a holomorphic matrix. 

Before discussing eq.(\ref{nontri}), we would like to fix the boundary condition at spatial 
infinity. 
It should be imposed that $V(T)=0$ at infinity, which implies that $T$ becomes proportional 
to a unitary matrix. 
We take the polar decomposition $V=UH$ and set $T_0=\sqrt{\zeta}\cdot 1$, where U is 
unitary and H is hermitian. 
Then the boundary condition means that $H\to 1$ at infinity. 

We define $R=V^\dagger V=H^2$. 
Then eq.(\ref{nontri}) can be rewritten in terms of $R$.
\begin{equation}
g^{a\bar{b}}\partial_{\bar{b}}(R^{-1}\partial_a R)=\zeta(R-1)
\end{equation}
The solutions of this equation in $N=1$ case are discussed in \cite{TS}. 
$U$ will be determined by requiring the regularity of the solutions. 

\vspace{3mm}

To the intersecting brane configuration including three kinds of D7-branes considered above, 
we can add D3-branes extended along (0789) without breaking more supersymmetries. 
Therefore eqs.(\ref{holo})-(\ref{nontri}) are expected to have the corresponding solutions 
and they will have nonzero third Chern character. 
From the boundary condition,
\begin{equation}
A_\mu \to iU\partial_\mu U^\dagger .
\end{equation}
The third Chern character can be calculated by the following expression.
\begin{equation}
\int_{{\bf R}^6} \mbox{tr}F^3 = -\frac i{10}\int_{\partial{\bf R}^6} \mbox{tr}UdU^\dagger 
                                dUdU^\dagger dUdU^\dagger
\end{equation}
Let us discuss the special case $N=4$, in which $U$ can be explicitly constructed,
\begin{equation}
U=\frac1r \gamma_m x_m \hspace{1cm} (m=1,\cdots,6),
\end{equation}
where $\gamma_m$ are the lower-left parts of the $SO(6)$ gamma matrices,
\begin{equation}
\Gamma_m=\left(
  \begin{array}{cc}
     & \gamma^\dagger_m \\ \gamma_m & 
  \end{array}
\right)
\end{equation}
and $r^2=(x^1)^2+\cdots+(x^6)^2$. 
This gives the asymptotic form of the gauge fields whose third Chern character is integrated 
to be unity, corresponding to the single D3-brane on the intersecting D7-branes. 

The asymptotic behavior of the tachyon field is also determined in this case as follows.
\begin{equation}
T \to \frac{\sqrt{\zeta}}r\gamma_m x_m 
\end{equation}
This is what has been discussed in the literature \cite{Ktheory}.

\vspace{1cm}

{\bf {\large 6. Discussions}}

\vspace{5mm}

In the previous section, we derived the equations (\ref{holo})-(\ref{nontri}) which 
described intersecting brane systems. 
Thus, by quantizing the collective coordinates of the solutions, the number of their 
bound states may be able to be determined, as will be explained below, 
if we assume their existence. 
In fact solving the BPS equations is not an easy task. 
However the number can be deduced from a few properties of the solutions and supersymmetry. 

The BPS equations are not scale invariant. 
Therefore all of the bosonic moduli of the solutions will correspond to the positions of 
them and the scale of them will be fixed. 
Since there remains some part of supersymmetry, the number $F$ of fermionic partners of the 
bosonic moduli will be a multiple of the number $B$ of the bosonic moduli. 
The low energy dynamics of the brane system will be described by quantum mechanics whose 
target space is the moduli space of the solutions, and in particular, the number of bound 
states will be determined by counting the number of ground states of quantum mechanics. 

Consider first D7-D7' system discussed in the previous section. 
In this case, it is expected that $B=4$ and $F=8$. 
This implies that the number of the bound states is $2^{\frac82}=16$. 
Since D7-D7' system is a T-dual of D4-D0 system, this result is appropriate \cite{Pol2}. 
Next consider D7-D7'-D7''-D3 system. 
In this case $B=6,F=6$ and the number of the bound states is $2^{\frac62}=8$. 
This is the value expected from the counting of a black hole entropy \cite{N=2} because this 
system is a T-dual of D4-D4'-D4''-D0 system. 

\vspace{3mm}

In our discussions, the tachyon potential $V(T)$ is determined by requiring the stability of 
the brane solutions. 
Thus we can calculate the tensions of the remaining branes as well as that of the original 
brane-antibrane system, and check whether our model reproduces the correct values. 
It is natural to set $g^{-2}=(2\pi \alpha')^2\tau_9$ and $\zeta=\frac1{2\alpha'}$, where 
$\tau_9$ is the tension of the BPS D9-brane. 
Then the tension of the original brane-antibrane system is
\begin{equation}
V(0)=\frac{\zeta^2}{2g^2}Tr(1)=\frac{\pi^2}4\cdot N\cdot 2\tau_9,
\end{equation}
and the tension of the resulting codimension-two brane is 
\begin{equation}
\frac{2\pi\zeta}{g^2}=\pi\tau_7. 
\end{equation}
These do not match the correct values. 
Probably this is because we have considered the action (\ref{action}), which includes only 
the leading terms in order of $\alpha'$. 
Therefore if one can construct an effective action containing all $\alpha'$ corrections, 
then we will be able to determine the form of the tachyon potential $V(T)$ by the same way 
as explained in this paper, 
and the model will provide the correct values for the D-brane tensions. 
The string field theory calculation \cite{SFT} seems to imply that the string loop 
corrections are not important, and our classical arguments will be enough to do the job. 

\vspace{2cm}

{\bf {\large Acknowledgments}}

\vspace{5mm}

I would like to thank H.Itoyama, T.Matsuo, K.Murakami, K.Hashimoto for valuable discussions. 
This work is supported in part by JSPS Reseach Fellowships.


\newpage

\begin{thebibliography}{99}

\bibitem{tachyon}M.Green, {\it POINT-LIKE STATES FOR TYPE 2b SUPERSTRINGS}, Phys. Lett. 
{\bf B329} (1994) 435, hep-th/9403040; \\
T.Banks, L.Susskind, {\it Brane - Anti-Brane Forces}, hep-th/9511194; \\
M.Green, M.Gutperle, {\it Light-cone supersymmetry and D-branes}, Nucl. Phys. 
{\bf B476} (1996) 484, hep-th/9604091; \\
G.Lifschytz, {\it Comparing D-branes to Black-branes}, Phys. Lett. {\bf B388} (1996) 720, 
hep-th/9604156; \\
V.Periwal, {\it Antibranes and crossing symmetry}, hep-th/9612215.

\bibitem{nonBPS}A.Sen, {\it Stable Non-BPS Bound States of BPS D-branes}, 
JHEP 9808 (1998) 010, hep-th/9805019; \\
A.Sen, {\it Tachyon Condensation on the Brane Antibrane System}, 
JHEP 9808 (1998) 012, hep-th/9805170; \\
A.Sen, {\it SO(32) Spinors of Type I and Other Solitons on Brane-Antibrane Pair}, 
JHEP 9809 (1998) 023, hep-th/9808141. 

\bibitem{SFT} A.Sen, B.Zwiebach, {\it Tachyon condensation in string field theory}, 
JHEP 0003 (2000) 002, hep-th/9912249 ; \\
W.Taylor, {\it D-brane effective field theory from string field theory}, hep-th/0001201; \\
N.Moeller, W.Taylor, {\it Level truncation and the tachyon in open bosonic string field 
theory}, hep-th/0002237; \\
J.Harvey, P.Kraus, {\it D-Branes as Unstable Lumps in Bosonic Open String Field Theory}, 
JHEP 0004 (2000) 012,hep-th/0002117; \\
R.de Mello Koch, A.Jevicki, M.Mihailescu, R.Tatar, {\it Lumps and P-branes in Open String 
Field Theory}, hep-th/0003031; \\
N.Berkovits, {\it The Tachyon Potential in Open Neveu-Schwarz String Field Theory}, 
JHEP 0004 (2000) 022, hep-th/0001084; \\
N.Berkovits, A.Sen, B.Zwiebach, {\it Tachyon Condensation in Superstring Field Theory}, 
hep-th/0002211; \\
P.De Smet, J.Raeymaekers, {\it Level Four Approximation to the Tachyon Potential in 
Superstring Field Theory}, hep-th/0003220; \\
N.Moeller, A.Sen, B.Zwiebach, {\it D-branes as Tachyon Lumps in String Field Theory}, 
hep-th/0005036; \\
A.Iqbal, A.Naqvi, {\it Tachyon Condensation on a non-BPS D-brane}, 
hep-th/0004015.

\bibitem{Ktheory}E.Witten, {\it D-Branes And K-Theory}, JHEP 9812 (1998) 019, 
hep-th/9810188; \\
P.Horava, {\it Type IIA D-Branes, K-Theory, and Matrix Theory}, Adv. 
Theor. Math. Phys. {\bf 2} (1999) 1373, hep-th/9812135.

\bibitem{TS}T.Suyama, {\it Description of Intersecting Branes via Tachyon Condensation}, 
hep-th/0006052, to appear in Phys. Lett. B.

\bibitem{vortex}H.Nielsen, P.Olesen, {\it Vortex Line Models for Dual Strings}, 
Nucl. Phys. {\bf B61} (1973) 45; \\
H.de Vega, F.Schaposnik, {\it Classical Vortex Solution of the Abelian Higgs Model}, 
Phys. Rev. {\bf D14} (1976) 1100; \\
E.Weinberg, {\it Multivortex Solutions of the Ginzburg-Landau Equations}, 
Phys. Rev. {\bf D19} (1979) 3008; \\
C.Taubes, {\it Arbitrary N-Vortex Solutions to the First Order Ginzburg-Landau Equations}, 
Comm. Math. Phys. {\bf 72} (1980) 277.

\bibitem{Pol2}J.Polchinski, {\it String Theory I,II}, Cambridge University Press.

\bibitem{N=2}J.Maldacena, {\it N=2 Extremal Black Holes and Intersecting Branes}, 
Phys. Lett. {\bf B403} (1997) 20, hep-th/9611163.


\end{thebibliography}
\end{document}